\documentclass[aps,twocolumn,showpacs,showkeys,nofootinbib]{revtex4-1}
\usepackage{epsfig}
\usepackage{amsmath}
\usepackage{amsfonts}
\usepackage{amssymb}
\usepackage{graphicx}
\usepackage{colordvi}
\usepackage{color}
\usepackage{dcolumn}
\usepackage{multirow}
\usepackage{hyperref}
\usepackage{epstopdf}
\usepackage{bm}
\usepackage{times}
\hypersetup{
  colorlinks=true,        
  linkcolor=blue,         
  citecolor=magenta,      
}

\begin{document}

\newcommand{\aea}{Astron. Astrophys.}
\newcommand{\apjl}{Astrophys. J. Lett.}
\newcommand{\cqg}{Class.  Quant. Grav.}
\newcommand{\grg}{Gen.  Rel. Grav.}
\newcommand{\jcap}{J. Cosmol. Astropart. Phys.}
\newcommand{\mnras}{Mon. Not. R. Astron. Soc.}
\newcommand{\plb}{Phys. Lett. B}

\title{Analysis of the Yukawa gravitational potential in $f(R)$ gravity I: semiclassical periastron advance.}

\author{Ivan De Martino}
\email{ivan.demartino@ehu.eus}
\affiliation{$^{1}$Department of Theoretical Physics and History of Science,  University of the Basque Country UPV/EHU, 
Faculty of Science and Technology, Barrio Sarriena s/n, 48940 Leioa, Spain}

\author{Ruth Lazkoz}
\email{ruth.lazkoz@ehu.eus}
\affiliation{$^{1}$Department of Theoretical Physics and History of Science,  University of the Basque Country UPV/EHU, 
Faculty of Science and Technology, Barrio Sarriena s/n, 48940 Leioa, Spain}

\author{Mariafelicia De Laurentis}
\email{laurentis@th.physik.uni-frankfurt.de}
\affiliation{Institute for Theoretical Physics, Goethe University,
Max-von-Laue-Str.~1, 60438 Frankfurt, Germany}
\affiliation{Dipartimento di Fisica "E. Pancini", Universit\'a di Napoli "Federico II", Compl. Univ. di Monte S. Angelo, Edificio G, Via Cinthia, I-80126, Napoli, Italy}

\affiliation{Lab.Theor.Cosmology,Tomsk State University of Control Systems and Radioelectronics(TUSUR), 634050 Tomsk, Russia}

\date{\today}

\begin{abstract}
The {\it concordance} cosmological model has been successfully tested over the last decades. 
Despite its successes, the fundamental nature of dark matter and dark energy is still unknown.
Modifications of the gravitational action have been proposed as an alternative to these dark components.
The straightforward modification of gravity is to generalize the action to a function, $f(R)$, of the scalar curvature. 
Thus one is able to describe the emergence and the evolution of the Large Scale Structure without 
any additional (unknown) dark component. In the weak field limit of the $f(R)$-gravity, 
a modified Newtonian gravitational potential arises. This gravitational potential accounts 
for an extra force, generally called fifth force,  that produces a precession of the orbital motion even in the 
classic mechanical approach. We have shown that the orbits in the modified potential 
can be written as Keplerian orbits under some conditions on the strength and scale length of this extra force. 
Nevertheless, we have  also shown that this extra term gives rise to the precession of the orbit. 
Thus, comparing our prediction with the measurements of the precession of some planetary motions, 
we have found that the strength of the fifth force must be in the range $[2.70-6.70]\times10^{-9}$ 
with the characteristic scale length to fix to the fiducial values of $\sim 5000$ AU.
\end{abstract}
\maketitle
\section{Introduction}
\label{uno}

Standard cosmology is entirely based on General Relativity. 
It is capable of explaining both the present period of accelerated expansion and 
the dynamics of self-gravitating systems resorting to Dark Energy and Dark Matter, respectively. 
The model has been confirmed by observations carried out over the last decades \cite{Planck16_13}.
The need of having recourse to Dark Matter to explain the dynamics of stellar clusters, 
galaxies, groups and clusters of galaxies, among others astrophysical objects, has been well known for many decades now.
Such systems show a deficit of mass when  the photometric and spectroscopic estimates are compared with the dynamical one. 
Early astronomical candidates proposed to solve this problem of {\em missing} mass  were MAssive Compact Halo Objects (MACHOs) and ReAlly Massive Baryon Objects (RAMBOs),
sub-luminous compact objects (or clusters of objects) like Black Holes and Neutron Stars that could
not have been observed due to several selection effects. Since the number of the observed sub-luminous 
objects was not enough to account for the {\em missing matter}, the idea that this matter was hidden in some exotic
particles, weakly interacting with ordinary matter, emerged. Many candidates have been proposed such as
Weakly Interacting Massive Particle (WIMP), axions, neutralino, Q-balls, gravitinos and Bose-Einstein condensate, 
among the others \cite{Bertone2005,Capolupo2010,Schive2014,demartino2017b,demartino2018,Lopes2018,Panotopoulos2018}, 
but there are no experimental evidences of their existence so far \cite{Feng2010}. 

An alternative approach is to modify Newton's law. Such a modification naturally
arises in the weak field limit of some modified gravity models \cite{Moffat2006, demartino2017, PhysRept,manos,sergei} 
that attempt to explain the nature of Dark Matter and Dark Energy as an effect 
of the space-time curvature. These theories predict the existence of massive gravitons 
that may carry the gravitational interaction over a certain scale depending by the mass of these particles \cite{Bogdanos:2009tn,felix,Bellucci:2008jt,graviton}. 
Thus, in their weak field limit, a Yukawa-like modification to Newton's law emerges. 
One of those models is $f(R)$-gravity where the Einstein-Hilbert action, 
which is linear in the Ricci scalar $R$, is replaced with a more general 
function of the curvature, $f(R)$. In its the weak field limit, the modified Newtonian potential
has the following functional form \cite{Annalen}:
\begin{equation}\label{eq:potyuk}
 \Phi(r) = -\frac{G M}{(1+\delta)r}(1+\delta e^{-r/\lambda}),
\end{equation}
where $M$ is the mass of the point-like source, $r$ the distance of a test particle ($m$) from the source, $G$ is Newton's constant,
$\delta$ is the strength of the Yukawa correction and the $\lambda$ represents the 
scale over which  the Yukawa-force acts.  Since $f(R)$-gravity is a fourth-order theory,  
the Yukawa scale length  arises from the extra degrees of freedom  
(in the general paradigm, a $(2k+2)$-order theory of gravity gives rise to $k$ extra gravitational scales \cite{Quandt1991}).
Both parameters are also related to the $f(R)$-Lagrangian as \cite{Annalen, PhysRept}:
\begin{align}
 \delta = f'_0 - 1, \qquad \lambda = \sqrt{-\frac{6f''_0}{f'_0}},
\end{align}
where  
\begin{equation}
f'_0=\frac{df(R)}{dR}\biggr|_{R=R_0}\,,  \qquad f''_0=\frac{d^2f(R)}{dR^2}\biggr|_{R=R_0}.
\end{equation}
Next, considering the field equations and trace of $f(R)$ gravity 
at the first order approximation in terms of the perturbations of the metric tensor, and choosing
a suitable transformation and a gauge condition, one can relate the massive states of the graviton
to the $f(R)$-Lagrangian and to the Yukawa-length:
\begin{align}
 m_g^2 \propto \frac{-f'_0}{f''_0} =\frac{2}{\lambda^2}.
\end{align}
Therefore, it is customary to identify the Yukawa-length with the  Compton wavelength 
of the massive graviton $\lambda_c = hc/m_g$. Thus, for example, 
we have $\lambda\sim10^{3}$ km with a mass of gravitons $m_g\sim 10^{-22}$ eV \cite{Lee2010,Abbott2017}. Therefore,
the effect of a modification of the Newtonian potential must naturally act at galactic and extragalactic scales,
where $f(R)$-gravity has been successfully tested \cite{demartino2014, demartino2015, demartino2016}.  
Nevertheless, smaller effects could be detected at shorter scales \cite{Talmadge1988} where
the strength of the Yukawa-correction has been bounded  using the Pioneer anomaly \cite{Anderson1998, Anderson2002} 
and S2 star orbits \cite{Borka20012,Borka20013, Zakharov2016, Hees2017, Zakharov2018,Iorio2005}. 
Obviously, the most interesting systems to test gravitational theories are binary systems 
composed by coalescing compact stars, such as neutron stars, white dwarfs and/or black holes 
\cite{deLa_deMa2014, deLa_deMa2015, LeeS2017}, but the study of stable orbits is equally important since 
it allows us to study possible variations of the gravitational interaction in the weak field limit. 
In the last decades, the orbital precession has been used to probe General Relativity \cite{Will2014, Iorio2009}, 
as well as 
to place bounds on “anti-gravity” due to the cosmological constant \cite{islam1983, Iorio2006, Sereno2006}, 
on forces proposed as alternatives to dark matter \cite{Gron1996, Khriplovich2006} and/or
induced from  extensions of General Relativity \cite{Capozziello2001, Moffat2006, Sanders2006, Battat2008, Nyambuya2010, Ozer2017, Liu2018}.

In this paper we  show, in a semi-classical approach, that the orbital motion under the 
modified gravitational potential in Eq. \eqref{eq:potyuk} can be traced back to a Keplerian orbit with
modified eccentricity, but with an orbital precession due to the Yukawa-term. 
We  consider two point-like masses orbiting around each other and we use a
Newtonian approach to compute the equation of the orbit, and a perturbative approach to compute the precession of the orbit. 
Finally, we use the current limits on the orbital precession of the planetary orbits to place a bound on
the  strength of the Yukawa-term.
The paper is organized as follows: in Sec. \ref{due} we introduce the equations of motion, in Sec. \ref{tre} we compute the 
equation of the orbits, in Sec. \ref{quattro} we compute analytically the precession effect due to the 
Yukawa potential, and we use current measurements of the orbital precession of 
Solar System's planets to bound the parameter $\delta$ in Eq. \eqref{eq:potyuk}.  
We consider, for each planet, a 3$\sigma$ interval around the best fit value of the precession, and we compute
the lower and an upper limit on $\delta$ so that 
the predicted precession relies in the observed interval. In Sec. \ref{cinque} we discuss some
consequences of our results. Finally, in Sec. \ref{sei} we give our conclusions.

\section{Newtonian approach to two body problem in Yukawa potential}
\label{due}
The starting point is the equation of motion of a massive point-like particle, $m$, in the gravitational
potential well generated by the particle $M$, and given in Eq. \eqref{eq:potyuk}. 
In polar coordinates $(r,\varphi)$ and with respect to the center of mass, the equations of motion read
\begin{align}
\label{eq:1} & \ddot{r} = -\nabla\Phi(r)\,,\\ 
\label{eq:2} & \frac{d}{dt}(r^2 \dot{\varphi}) = 0 \,, 
\end{align}
and the total energy of the system can be written as \cite{deLa2011}
\begin{equation}\label{eq:energy}
 E_T = \frac{1}{2}\mu(\dot{r}^2 + r^2\dot{\varphi}^2)-\frac{Gm M}{(1+\delta)r}(1+\delta e^{-r/\lambda}),
\end{equation}
where 
$\displaystyle{\mu=\frac{mM}{m+M}}$ is the reduced mass, and $\Phi(r)$ is the modified gravitational potential of Eq. \eqref{eq:potyuk}.
Using the conservation of the angular momentum $L$ expressed in Eq. \eqref{eq:2}, it is straightforward 
to recast the total energy as a function of  the radial coordinate:
\begin{equation}\label{eq:energy2}
 E_T = \frac{1}{2}\mu\dot{r}^2 +  \frac{L^2}{2\mu r^2} -\frac{Gm M}{(1+\delta)}\frac{(1+\delta e^{-r/\lambda})}{r}.
\end{equation}

Eq. \eqref{eq:energy2} is the only one needed to compute the equation of motion for an unperturbed orbit.
Nevertheless, we can learn more about the orbits by defining an effective potential as
\begin{equation}\label{eq:Veff}
 V_{\rm eff}(r) = \frac{L^2}{2\mu r^2} - \frac{Gm M}{(1+\delta) r} - Gm M \frac{\delta}{(1+\delta) r} e^{-r/\lambda}.
\end{equation}
Here, the first term accounts for the repulsive force associated to the angular momentum, the second term represents the gravitational
attraction, and the third term can be interpreted as an additional force due to the Yukawa-like term in the gravitational
potential acting on the particle. The effective potential demands other considerations: first, one needs $\delta \neq -1$ 
in order to avoid a singularity in the second and third terms; second, if $\delta$ assumes negative values, 
the second term stays attractive as far as the condition $ \delta > -1 $ is satisfied, and the last term becomes repulsive;
third, the condition $\delta <-1$ makes the second term repulsive, rendering the third term attractive; 
fourth, if $\delta >0$ then both second and third terms are attractive.

\begin{figure*}
\centering
\includegraphics[width=8.6cm]{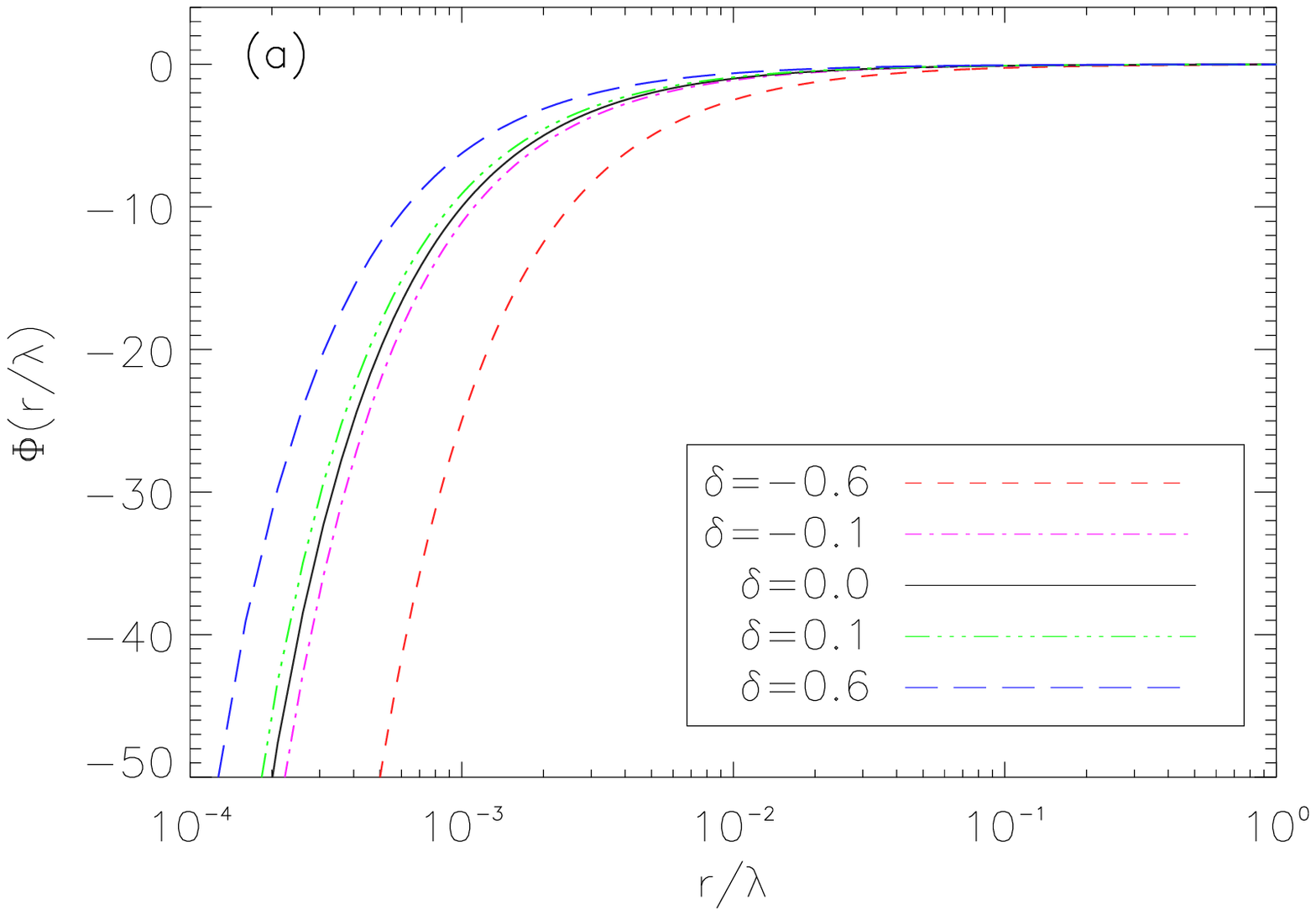}
\includegraphics[width=8.6cm]{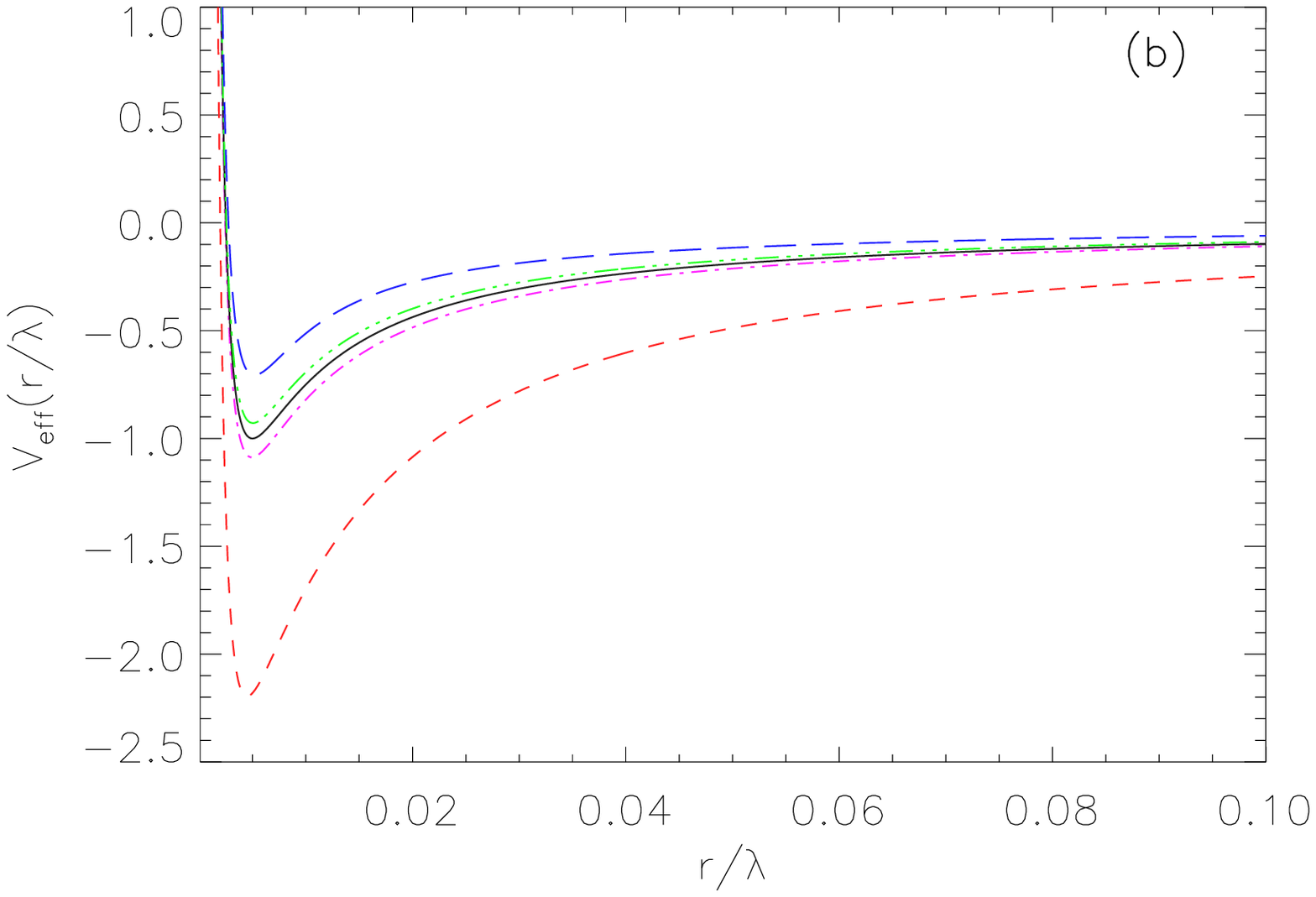}
\caption{Modified gravitational and effective potentials as a function of the distance 
from the gravitational source $M$. Solid black lines indicate the Newtonian case ($\delta=0$), 
dashed colored lines depict the corrections. }\label{fig1}
\end{figure*}

For illustration, in Fig. \ref{fig1}(a) and (b) we plot the potential  and the effective potential as a function of $r/\lambda$ 
showing their dependence on the strength of the Yukawa term. Notice that the  minimum of the effective potential 
depends on the strength parameter $\delta$ of the Yukawa-term (Fig. \ref{fig1}(b)). As expected, a negative value of $\delta$ 
makes the potential well deeper as compared to the Newtonian case ($\delta=0$), while a positive value makes it flatter. This can be 
understood looking at Eq. \eqref{eq:potyuk}, for $-1<\delta<0$ the effective mass $M'= M/(1+\delta)$ becomes larger, while for 
$\delta>0$ it becomes smaller than the ``Newtonian  mass" $M$.

Differentiating with respect to the radial coordinate  and looking for the minimum, one finds the condition: 
\begin{equation}
\frac{d V_{\rm eff}(r)}{dr} =0 \Rightarrow \frac{L^2}{\mu r} = \frac{G m M \left(\delta  e^{-\frac{r}{\lambda}}+1\right)}{(\delta +1)}+\frac{\delta  G m M e^{-\frac{r}{\lambda}}}{(\delta +1) \lambda}r\,.
\end{equation}
The second derivative and the previous condition on the angular momentum leads to obtain the following
expression
\begin{equation}
\frac{d^2V_{\rm eff}(r)}{dr^2}= \frac{G m M  e^{-\frac{r}{\lambda}}}{(\delta +1)r^3}\biggl[ \delta(-\lambda^{-2}r^2
+ \lambda^{-1}r +1) +  e^{\frac{r}{\lambda}} \biggr].
\end{equation}
A minimum in the effective potential exists if the following condition is satisfied
\begin{equation}\label{eq:condition1}
 g(x)\equiv \delta(-x^2 + x +1) +  e^x>0\,,
\end{equation}
where we have defined $x\equiv r/\lambda$. Eq. \eqref{eq:condition1}
is satisfied in the following cases: (i)  $\delta>-1$  for  $x\rightarrow0$, (ii)
$\delta>-e$  for  $x\rightarrow1$, and (iii) $\forall\,\delta$ in the limit $x\rightarrow\infty$.
Let us notice that the first case, meaning $r\ll \lambda$, is the common configuration of an astrophysical system with its dynamics 
happening at scales much lower than the Compton wavelength of the massive graviton, such as planetary motion around the Sun.
On the contrary, the second case ($r\sim\lambda$) represents systems such as S-Stars around 
the Galactic center, with their dynamics happening at scales of the order of parsecs. Finally, the last case ($r\gg \lambda$) 
can be associated to the extragalactic and cosmological scales.
Since we are interested in studying systems on distance scales much smaller than the Compton scale of a massive graviton, 
the exponential term in previous equations Taylor expanded as 
\begin{equation}\label{eq:approx_exp}
 e^{\pm x} \approx 1 \pm x  + \frac{x^2}{2} + \mathcal{O}\bigl(x^3\bigr).
\end{equation}
When replacing Eq. \eqref{eq:approx_exp} in to Eq. \eqref{eq:potyuk}, the first term gives the Newtonian force,
the second term induces a shift in the energy of the system, and the third
term gives rise to a constant radial acceleration (often called fifth force) that can be written as follows
\begin{equation}
 a_{\rm corr} = -\frac{a^*\delta}{2(1+\delta)}\frac{r^{*2}}{\lambda^2},
\end{equation}
where $a^*$ is the Newtonian acceleration of an object at distance $r^*$. As an example, this correction can be applied
to the Pioneer anomaly, thus obtaining a strength $|\delta|\leq 1.7\times10^{-4}$ at $\lambda\sim200$ AU \cite{Anderson1998, Anderson2002}. 
It is important to remark that the approximation in Eq. \eqref{eq:approx_exp} is valid only for dynamics at the scale of planetary 
systems or stars with orbits having their semi-major axis much smaller than $\lambda$. In contrast, to study the dynamics 
of systems on larger scales, 
one cannot use the approximation in Eq. \eqref{eq:approx_exp} but rather, the equations of motion must be integrated numerically. 

Let us analyze the condition for the existence a minimum in the effective potential at both $\mathcal{O}(x^{2})$ 
and  $\mathcal{O}(x^{3})$ orders:
\begin{description}
 \item[$\mathcal{O}(x^{2})$ order]  at this order of approximation, the effective
      potential becomes
      \begin{equation}
      V_{eff}(r) =  \frac{L^2}{2 \mu  r^2} - \frac{G m M}{r}+\frac{\delta  G m M}{(\delta +1) \lambda},
      \end{equation}
      and we find the minimum at the radius 
      \begin{equation}\label{eq:rmin_order1}
       r_{min} = \frac{L^2}{2\mu G m M},
      \end{equation}
      which is the same as the one of Newtonian gravity (as expected), while the effective potential at the minimum is shifted
      with respect to the Newtonian one
      \begin{eqnarray}\label{eq:veffmin_order1}
       &&V_{eff, min}= - \frac{1}{2} G m M \left(\frac{G \mu  m M}{L^2} -\frac{2 \delta }{(\delta +1)\lambda}\right)\,,\nonumber\\
      \end{eqnarray}

 \item[$\mathcal{O}(x^{3})$ order] the effective potential can be recast as
      \begin{eqnarray}\label{eq:veff_order2}
      &&V_{eff}(r) =  \frac{L^2}{2 \mu  r^2} - \frac{G m M}{r}+\frac{\delta  G m M}{(\delta +1) \lambda} -\frac{\delta  G m M r}{2 (\delta +1) \lambda^2}\,.\nonumber\\
      \end{eqnarray}
      Since we are looking for a strength force in the regime $\delta\ll1$, thus meaning a small deviation from Newtonian dynamic,
      the shift in $r_{min}$ is absolutely negligible. 
      Thus, replacing Eq. \eqref{eq:rmin_order1} in to Eq. \eqref{eq:veff_order2} we get
      \begin{eqnarray}\label{eq:veffmin_order2}
       V_{eff, min}& =&  - \frac{G m M }{2} \left(\frac{G \mu  m M}{L^2} -\frac{2 \delta }{(\delta+1)\lambda}\right) \nonumber\\ 
                     &-&\frac{L^2 \delta }{2 (1+\delta ) \lambda ^2 \mu }\,.
      \end{eqnarray}
            
\end{description}

Therefore, at both $\mathcal{O}(\lambda^{-2})$ and  $\mathcal{O}(\lambda^{-3})$ orders, the minimum of the effective potential 
always exists and it is located at the same radius ($r_{min}$) than in the Newtonian case. Finally, Eqs. \eqref{eq:veffmin_order1}
and \eqref{eq:veffmin_order2} show that the minimum of the effective potential 
is shifted as qualitatively explained above and shown in Fig. \ref{fig1}b.

\section{Equation of the orbits}
\label{tre}
Hereby, we  compute the equation of the closed orbit 
in both $\mathcal{O}(x^{2})$ and $\mathcal{O}(x^{3})$ approximations, and we  show that, under
some conditions on the eccentricity and the position of latus rectum, the orbit can be recast into the usual Keplerian form, where
the correction due to the Yukawa-term getting hidden into the orbital parameters. We work in the regime $r\ll\lambda$  in order to
replace the exponential term in Eq. \eqref{eq:energy2} with Eq. \eqref{eq:approx_exp}.

\subsection{Approximation at $\mathcal{O}(x^{2})$-order}
\label{treA}
\begin{figure}
\centering
\includegraphics[width=8.6cm]{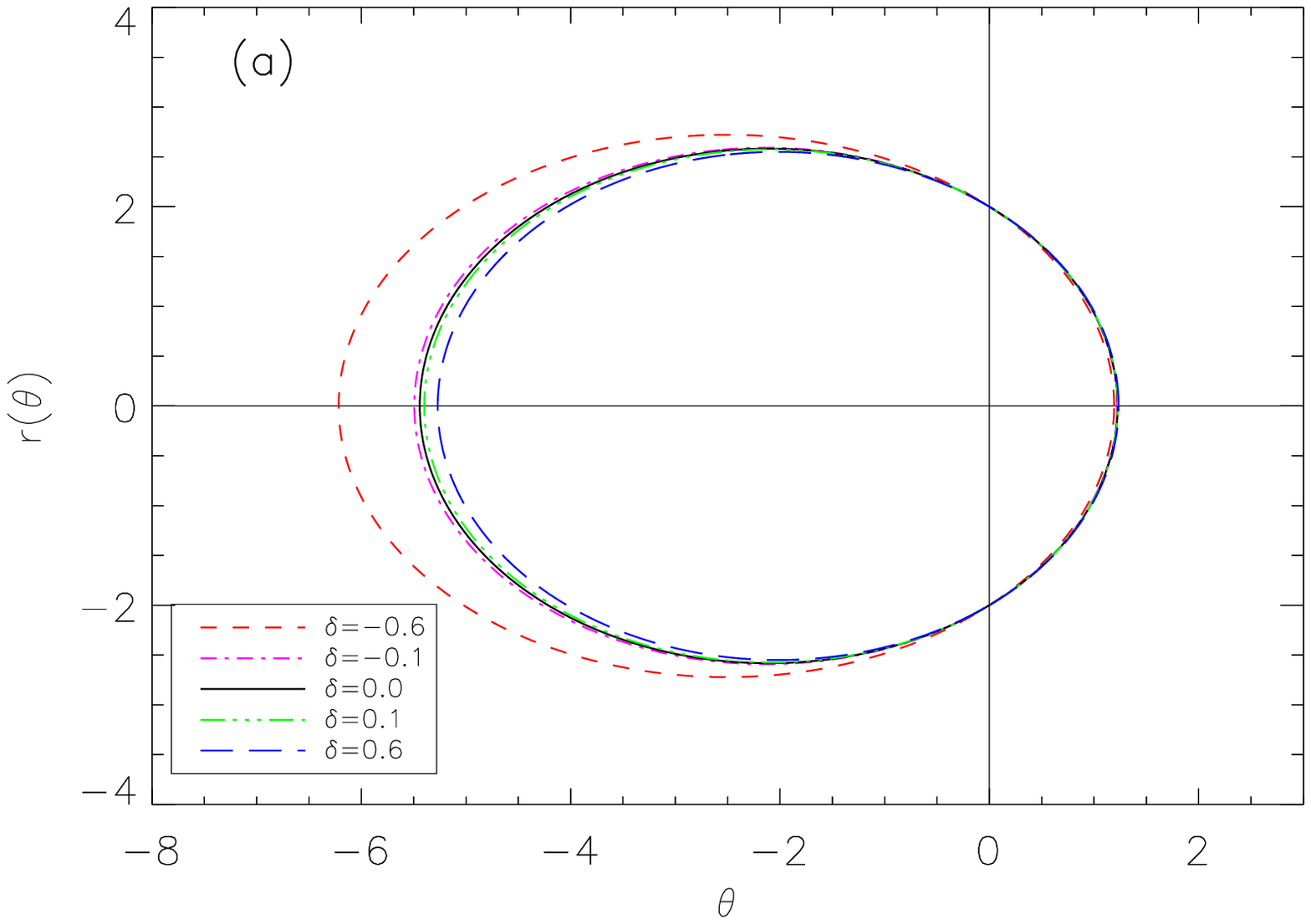}
\includegraphics[width=8.6cm]{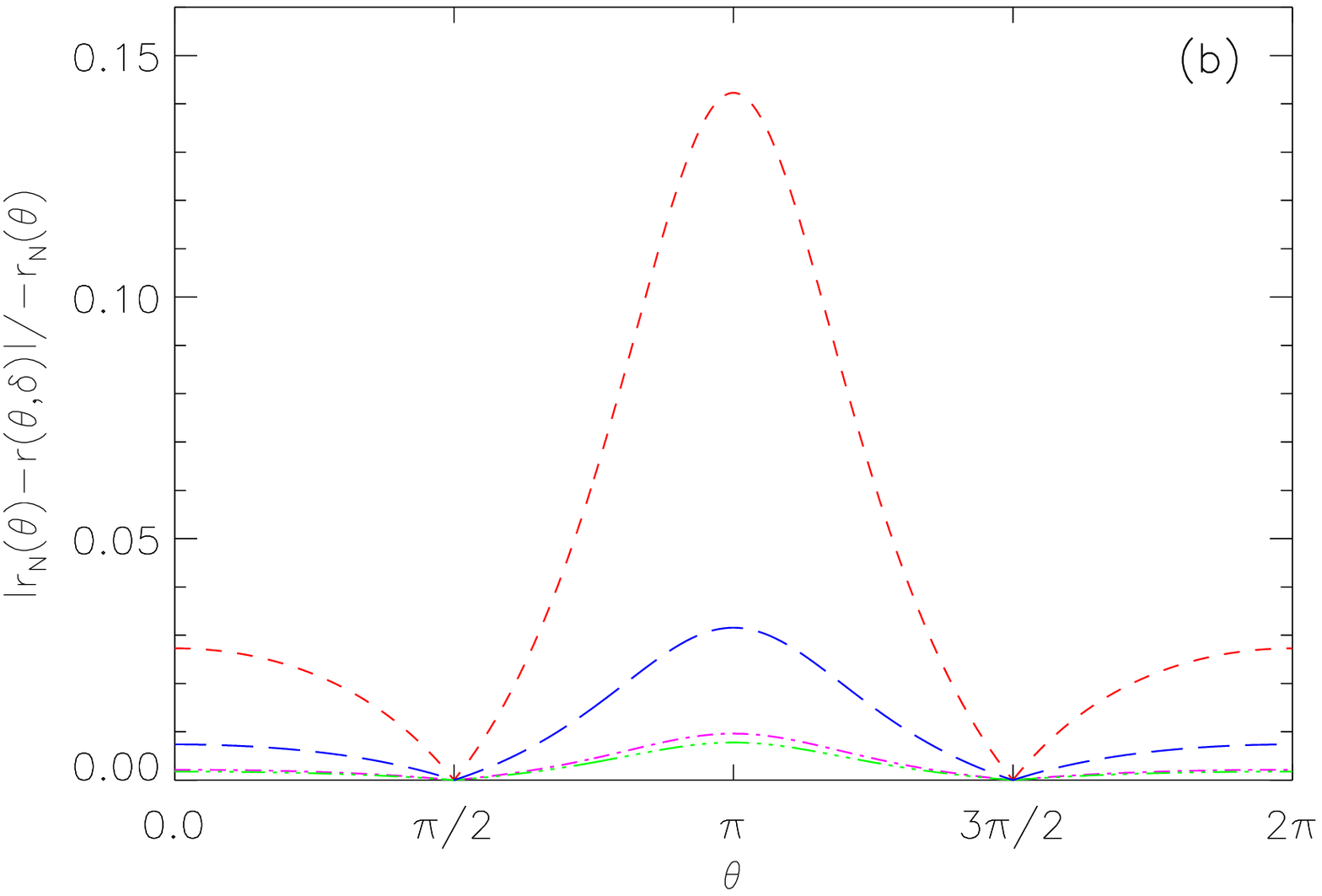}
\caption{Illustration of the effect of the modified gravitational potential on the orbital parameters. 
Panel (a) shows the orbits for different values of $\delta$. 
The angular momentum and the total energy are set to those values that give rise to an elliptical orbit with 
eccentricity $\epsilon=0.5$ in Newtonian mechanics ($\delta=0$) showing that such an orbital solution would show
a difference in the eccentricity when the Yukawa term is taken in to account. 
Panel (b) shows the relative difference with the Newtonian mechanics along the orbit.}\label{fig2}
\end{figure}

To compute the equation of the orbit we rewrite the radial component of the velocity as
\begin{equation}
\dot{r}=-\dfrac{L}{\mu}\dfrac{d}{d\varphi}\dfrac{1}{r}\,,
\end{equation}
then, at second order in the approximation of the Yukawa-term, the total energy of the system can be recast as
\begin{equation}\label{eq:energy3}
 E_T = \frac{L^2}{2\mu}\left(\dfrac{d}{d\varphi}\dfrac{1}{r}\right)^2 + \frac{L^2}{2\mu r^2}  -\frac{Gm M}{r} + \frac{Gm M \delta}{(1+\delta)\lambda}.
\end{equation}
From the previous equation we can obtain
the following differential equation
\begin{equation}\label{eq:energy4}
 u'^2+u^2 - 2\beta_0 u = \beta_1,
\end{equation}
where $u\equiv1/r$, $u'=du/d\varphi$ and
\begin{equation}\label{eq:betas01}
\gamma = G m M; \qquad  \beta_0=\frac{\mu\gamma}{L^2}; \qquad \beta_1=  \frac{2\mu E_T}{L^2}- \frac{2\mu\gamma}{L^2\lambda}\frac{\delta}{1+\delta}.
\end{equation}
Differentiating Eq. \eqref{eq:energy4}, we get
\begin{equation}\label{eq:energy5}
 u'\biggr(u''+u- \beta_0\biggr) = 0.
\end{equation}
As we are looking for a Keplerian solution, we make the following {\em ansatz}:
\begin{equation}\label{eq:orbit0}
 u\equiv\frac{1}{r}=\frac{1}{l}(1+\epsilon\cos\varphi),
\end{equation}
where $l$ is the {\em latus rectum} and $\epsilon$ is the eccentricity. Therefore, inserting the Eq. \eqref{eq:orbit0}
in Eq. \eqref{eq:energy5}, we obtain the following condition for the {\em latus rectum}:
\begin{equation}\label{eq:orbit1}
 l=\frac{1}{\beta_0}.
\end{equation}
Then, we substitute Eqs. \eqref{eq:orbit0} into Eq. \eqref{eq:energy4} thus obtaining the following expression
for the eccentricity:
\begin{equation}\label{eq:orbit2}
 \epsilon^2=1 + l^2 \beta_1,
\end{equation}
that in terms of energy of the system is
\begin{equation}\label{eq:orbit3}
 \epsilon^2 = 1 - \frac{2 L^2}{\mu\gamma} \frac{\delta}{(1+\delta)\lambda} + \frac{2 E_T L^2 \mu}{\mu^2\gamma^2},
\end{equation}
which for $\delta=0$ gets reduced to the Newtonian value:
\begin{equation}\label{eq:orbit3_1}
 \epsilon^2 = 1  + \frac{2 E_T L^2 \mu}{\mu^2\gamma^2}.
\end{equation}
This shift is clearly not testable with observations given that we measure the orbital parameters, whereas
the total energy is a theory dependent parameter. Nevertheless, looking at Eq. \eqref{eq:orbit3}, 
it is straightforward to understand that, if the total energy and angular 
momentum are fixed then they correspond to an orbital motion with an eccentricity  that 
would vary depending on the strength of the Yukawa correction, as shown in Figure \ref{fig2}.
\subsection{Approximation at  $\mathcal{O}(x^{3})$ order}
\label{treB}
Approximating the Yukawa-term at  third order, the differential equation \eqref{eq:energy4} becomes
\begin{equation}\label{eq:energy6}
 u'^2+u^2- 2\beta_0u - \beta_2\frac{1}{u} = \beta_1,
\end{equation}
where $\beta_0$ and $\beta_1$ are given in Eq. \eqref{eq:betas01}, and
\begin{equation}\label{eq:beta2}
 \beta_2=\frac{\mu\gamma\delta}{2 L^2\lambda(1+\delta)}.
\end{equation}
By taking the derivative of Eq. \eqref{eq:energy6} we obtain
\begin{equation}\label{eq:energy7}
 u'\left(u''+u +\frac{\beta_2}{u^2}-\beta_0\right)=0\,.
\end{equation}

\begin{figure}[t]
\centering
\includegraphics[width=8.6cm]{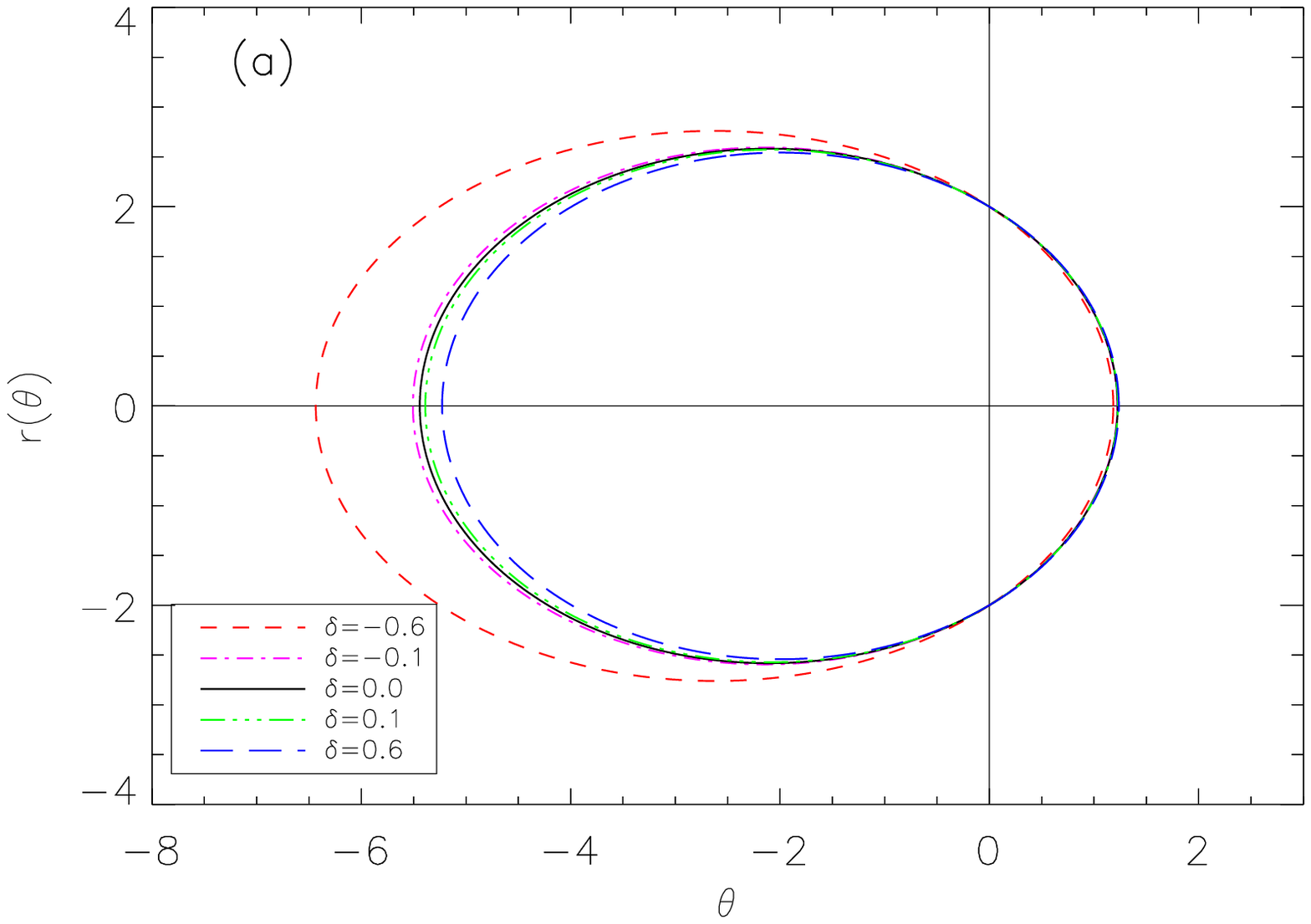}
\includegraphics[width=8.6cm]{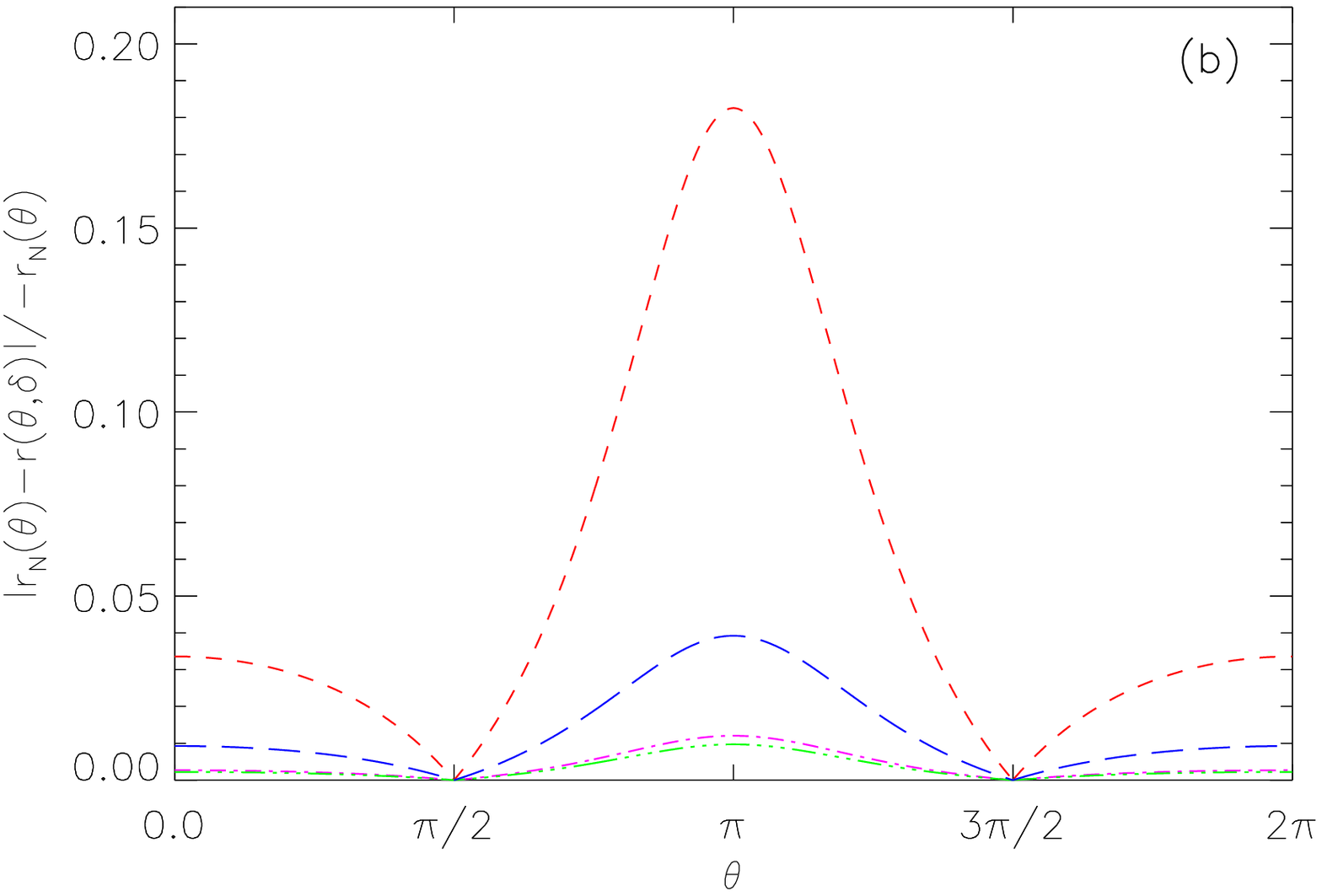}
\caption{The plots and panels replicate the ones in Fig. \ref{fig2} for the $\mathcal{O}(\lambda^{-3})$ approximation order.}\label{fig3}
\end{figure}

Let us introduce Eq. \eqref{eq:orbit0} into Eq. \eqref{eq:energy7} and evaluate the expression at $\varphi=[0; \pi]$, 
which respectively correspond
to the minimum and maximum distance between the two masses. Thus,  we obtain two conditions:
\begin{align}
& \label{eq:orbit4}  (1 - l \beta_0) \epsilon^2 + 2 (1 - l \beta_0) \epsilon - l \beta_0 + l^3  \beta_2 +1 =0\,, \\
& \label{eq:orbit5} (1 - l \beta_0) \epsilon^2 - 2 (1 - l \beta_0) \epsilon - l \beta_0 + l^3  \beta_2 +1 =0\,.
\end{align}
Subtracting Eqs. \eqref{eq:orbit4} and \eqref{eq:orbit5} we obtain the latus rectum which turns out to have the same
expression as in Eq. \eqref{eq:orbit1}. Finally,
introducing Eq. \eqref{eq:orbit0} in Eq. \eqref{eq:energy6} and evaluating it, once again, at $\varphi=[0; \pi]$ 
we obtain the following condition for the eccentricity
\begin{equation}
 \epsilon^2= 1+ l^2\beta_1-4 \beta_2.
\end{equation}
Let us note that the previous expression reduces to Eq. \eqref{eq:orbit2} when $\beta_2=0$, and thus to the Newtonian
value when $\delta=0$. The previous equation can be straightforwardly recast in terms of energy of the system as
\begin{equation}\label{eq:orbit6}
 \epsilon^2 = 1 + \frac{2 E_T L^2 \mu}{\mu^2\gamma^2} - \frac{2 L^2}{\mu\gamma} \frac{\delta}{(1+\delta)\lambda} - \frac{2\mu\gamma\delta}{L^2\lambda(1+\delta)}.
\end{equation}
The $\mathcal{O}(x^{3})$ order the shift is  larger than at $\mathcal{O}(x^{2})$ order,
and the difference due to the order of approximation is not negligible (see Fig. \ref{fig3}).

\section{Precession in Yukawa potential}
\label{quattro}
To compute analytically the periastron advance due to the Yukawa-like term in the gravitational potential, we study small perturbations
to the circular orbit. Thus, let us recast the total energy  as 
\begin{equation}\label{eq:prec1}
 u'^2+u^2+ \frac{g(u)}{L^2} = \frac{2\mu E_T}{L^2} - \frac{2\mu\gamma}{L^2\lambda}\frac{\delta}{1+\delta},
\end{equation}
where $ g(u)$ account for the gravitational interaction. Let us impose a close orbit defined by a minimum and a maximum distance
from the center: $r_-|_{\varphi=0} =a(1-\epsilon) $ and $r_+|_{\varphi=\pi} =a(1+\epsilon)$, respectively. 
Here $a$ is the semi-major axis of the orbit. Thus, those correspond to $u_0=1/r_-$ and $u_1=1/r_+$. Being $u'|_{u=u_0}=u'|_{u=u_1}=0$,
the Eq. \eqref{eq:prec1} gives rise to the following two conditions
\begin{align}
 & u_0^2+ \frac{g(u_0)}{L^2} = \frac{2\mu E_T}{L^2} - \frac{2\mu\gamma}{L^2\lambda}\frac{\delta}{1+\delta}\,,\\
 & u_1^2+ \frac{g(u_1)}{L^2} = \frac{2\mu E_T}{L^2} - \frac{2\mu\gamma}{L^2\lambda}\frac{\delta}{1+\delta}\,,
\end{align}
from which one obtains 
\begin{align}
 &  L^2 = \frac{g(u_0) - g(u_1)}{u_1^2 -  u_0^2}\,,\\
 &  E_T =  \frac{u_1^2g(u_0) - u_0^2g(u_1)}{2\mu(u_1^2 -  u_0^2)} + \frac{\mu\gamma\delta}{\mu(1+\delta)\lambda}\,.
\end{align}
Then, the differential equation \eqref{eq:prec1} becomes
\begin{equation}\label{eq:prec2}
 u' = \sqrt{G(u_0,u_1,u)}\,,
\end{equation}
where $G(u_0,u_1,u)$ is 
\begin{widetext}
\begin{eqnarray}
G(u_0,u_1,u)=\frac{g(u_0)(u_1^2-u^2) + g(u_1)(u^2-u_0^2)-(b^2 -  u_0^2)g(u)}{g(u_0) - g(u_1)}\, .
\end{eqnarray}
\end{widetext}
We can find the amount of angle  required to pass from $r_-$ to $r_+$ by integrating equation \eqref{eq:prec2}:
\begin{equation}\label{eq:prec_angle}
\varphi(r_+)-\varphi(r_-) = \int_{u_0}^{u_1} G(u_0,u_1,u)^{-1/2}du\,.
\end{equation}
Hence the particle will move from $r_-$ to $r_+$ 
and back every time $\varphi\rightarrow\varphi+2\pi$,
thus $r(\varphi)$ is periodic with period $2\pi$. Therefore, 
the precession for each revolution is
\begin{equation}
 \omega = 2|\varphi(r_+)-\varphi(r_-)| - 2\pi.
\end{equation}

In the case of approximating the exponential term at  $\mathcal{O}(x^{2})$ order, the function $g(u)$ 
only depends by a Newtonian term:
\begin{equation}\label{eq:prec3}
 g(u) = -2\mu\gamma u = 2 \mu \Phi_N(1/u),
\end{equation}
where $\Phi_N(1/u)$ is the classical Newtonian potential. 
Thus, the precession does not exist as expected for the Newtonian potential.

Nevertheless, when approximating the exponential term at  $\mathcal{O}(\lambda^{-3})$ order we have 
\begin{equation}\label{eq:prec4}
 g(u) = 2 \mu \Phi_N(1/u) - \frac{\mu\gamma\delta}{\lambda(1+\delta)} \frac{1}{u}.
\end{equation}
In order to solve the integral in Eq. \eqref{eq:prec3} we perform a change of variables
\begin{equation}
 u_1 = u_0+\eta; \qquad u= u_0+\eta\upsilon\,,
\end{equation}
with $0<\upsilon<1$. Then, the Eq. \eqref{eq:prec_angle} can be recast as
\begin{equation}\label{eq:prec7}
 \Delta \varphi \equiv \varphi(r_+)-\varphi(r_-) =  \eta \int_0^1 g(u_0,u_0+\eta,\lambda, \delta, u_0+\eta\upsilon)d\upsilon\,,
\end{equation}
where
\begin{equation}
 g(u_0,u_0+\eta,\lambda, \delta, u_0+\eta\upsilon) = \frac{1}{\sqrt{G(u_0,u_0+\eta,\lambda, \delta, u_0+\eta\upsilon)}}.
\end{equation}

\begin{figure*}
\begin{center}
\includegraphics[width=8.9cm]{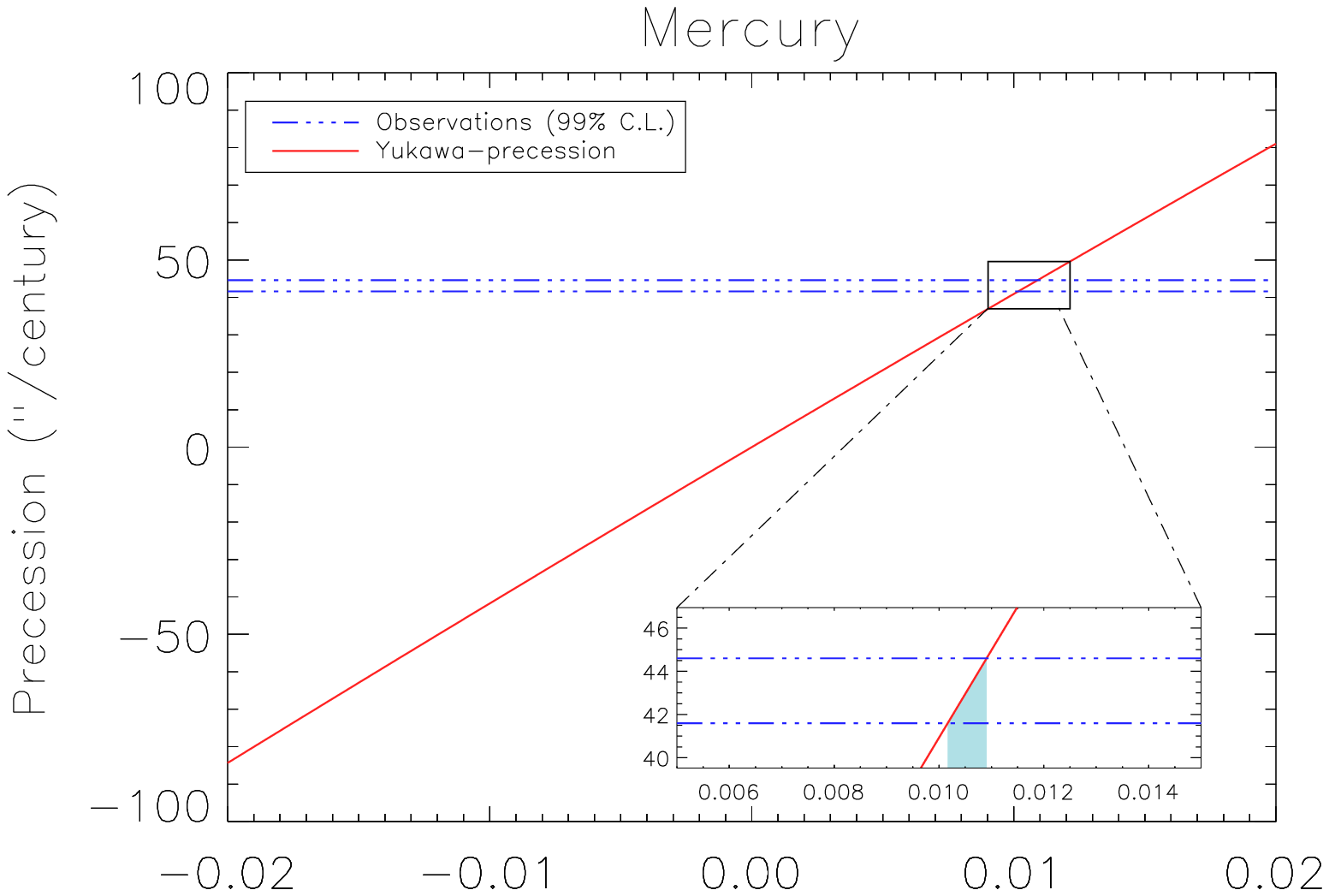}
\includegraphics[width=8.9cm]{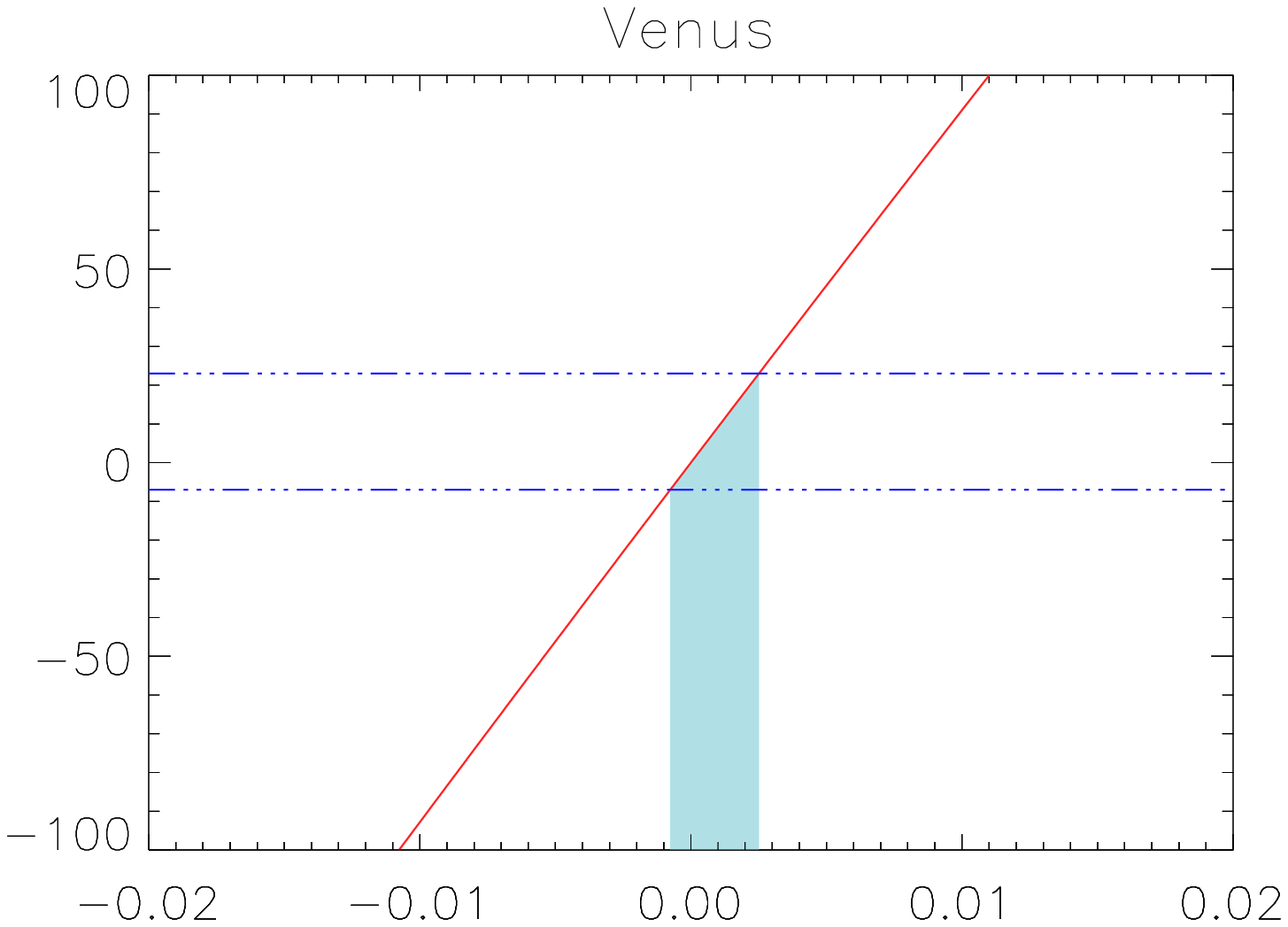}\\
\includegraphics[width=8.9cm]{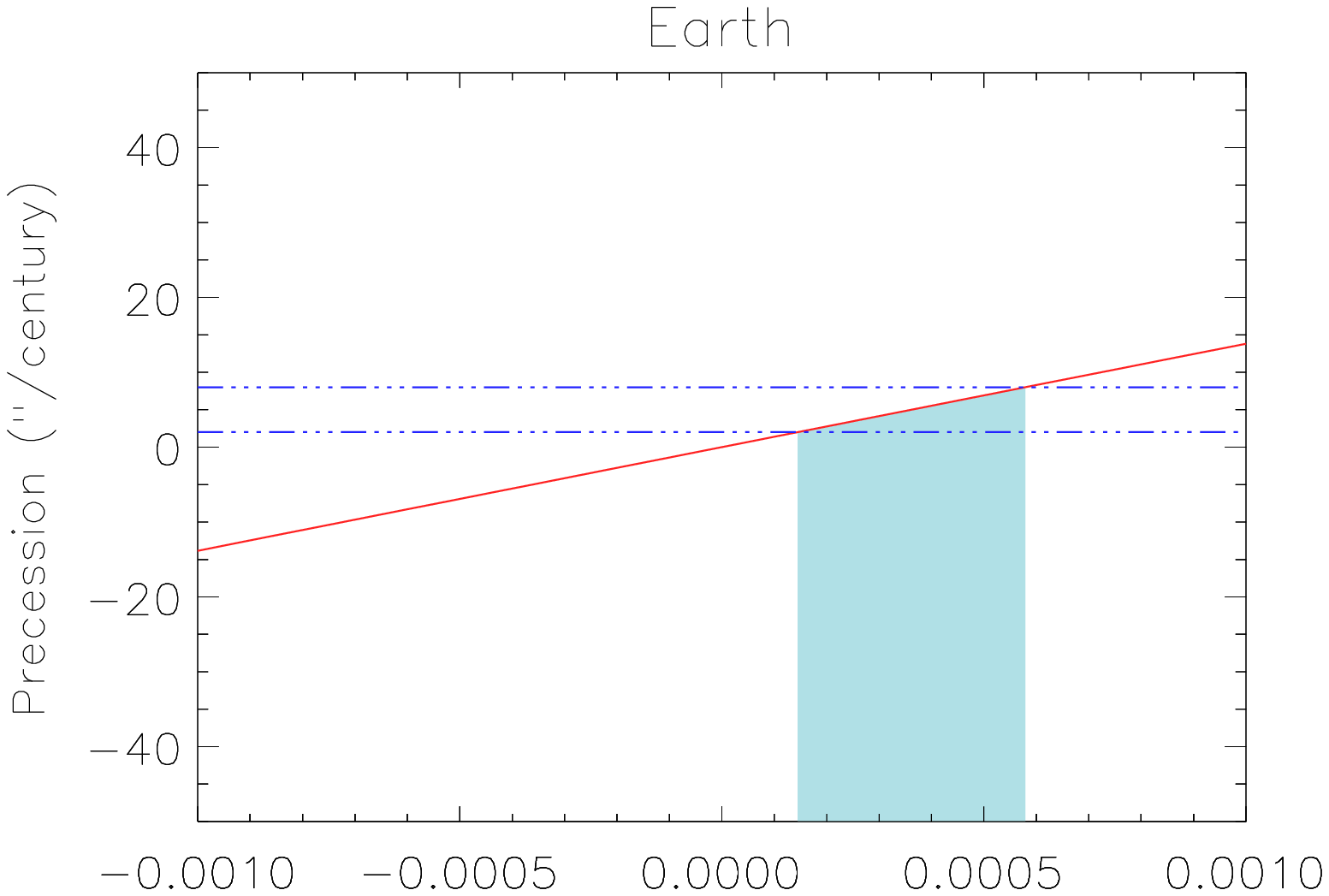}
\includegraphics[width=8.9cm]{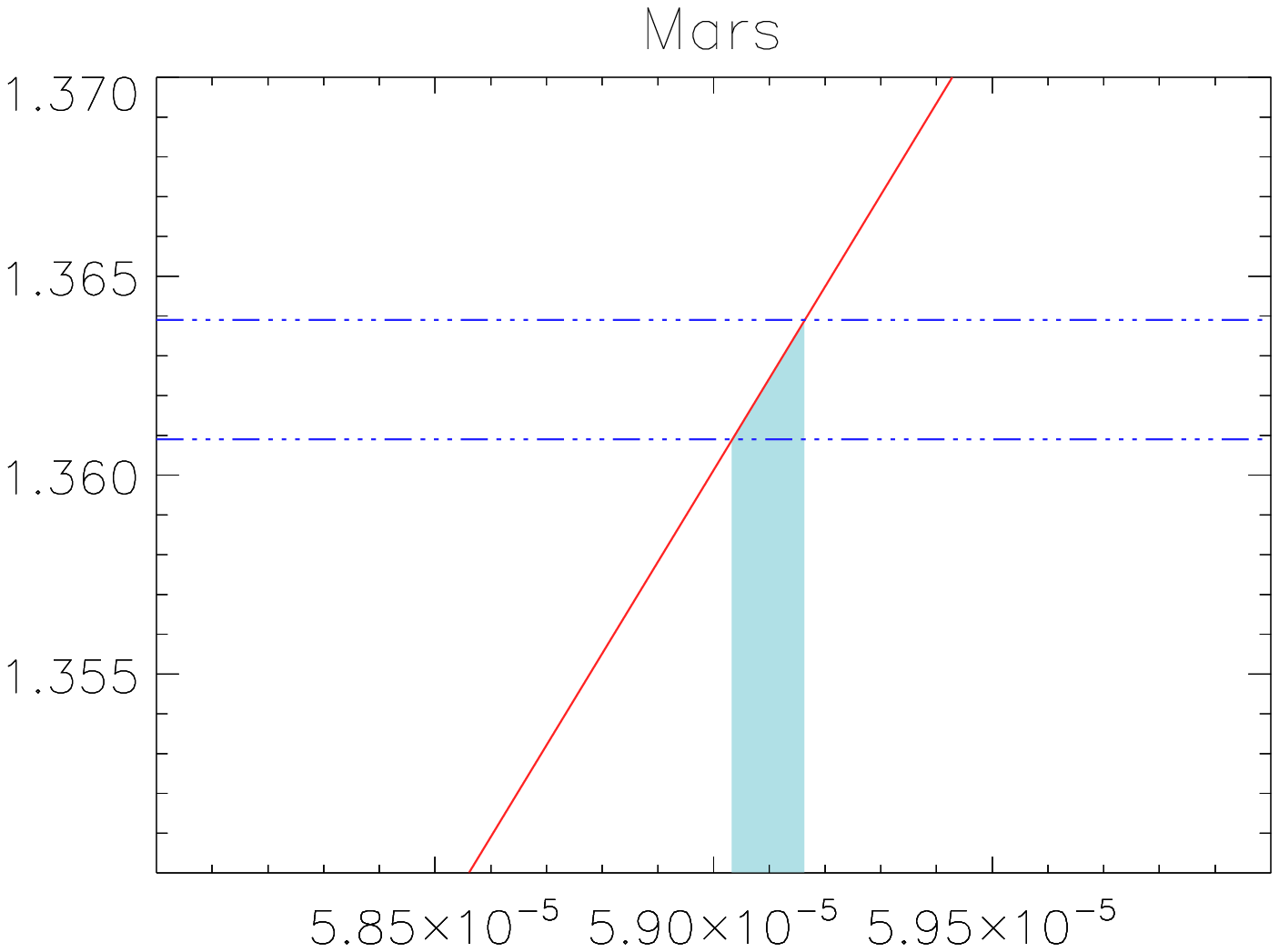}\\
\includegraphics[width=8.9cm]{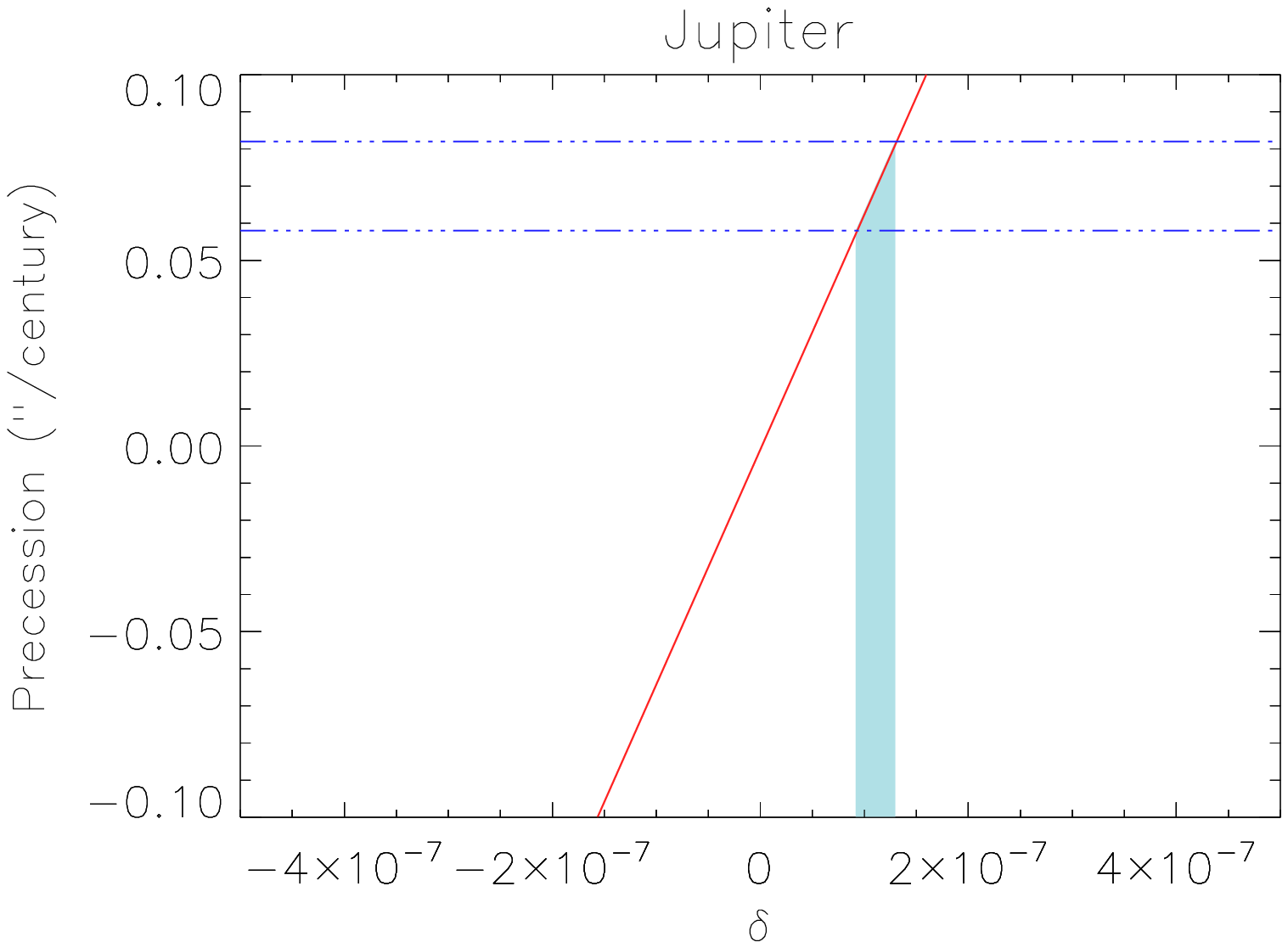}
\includegraphics[width=8.9cm]{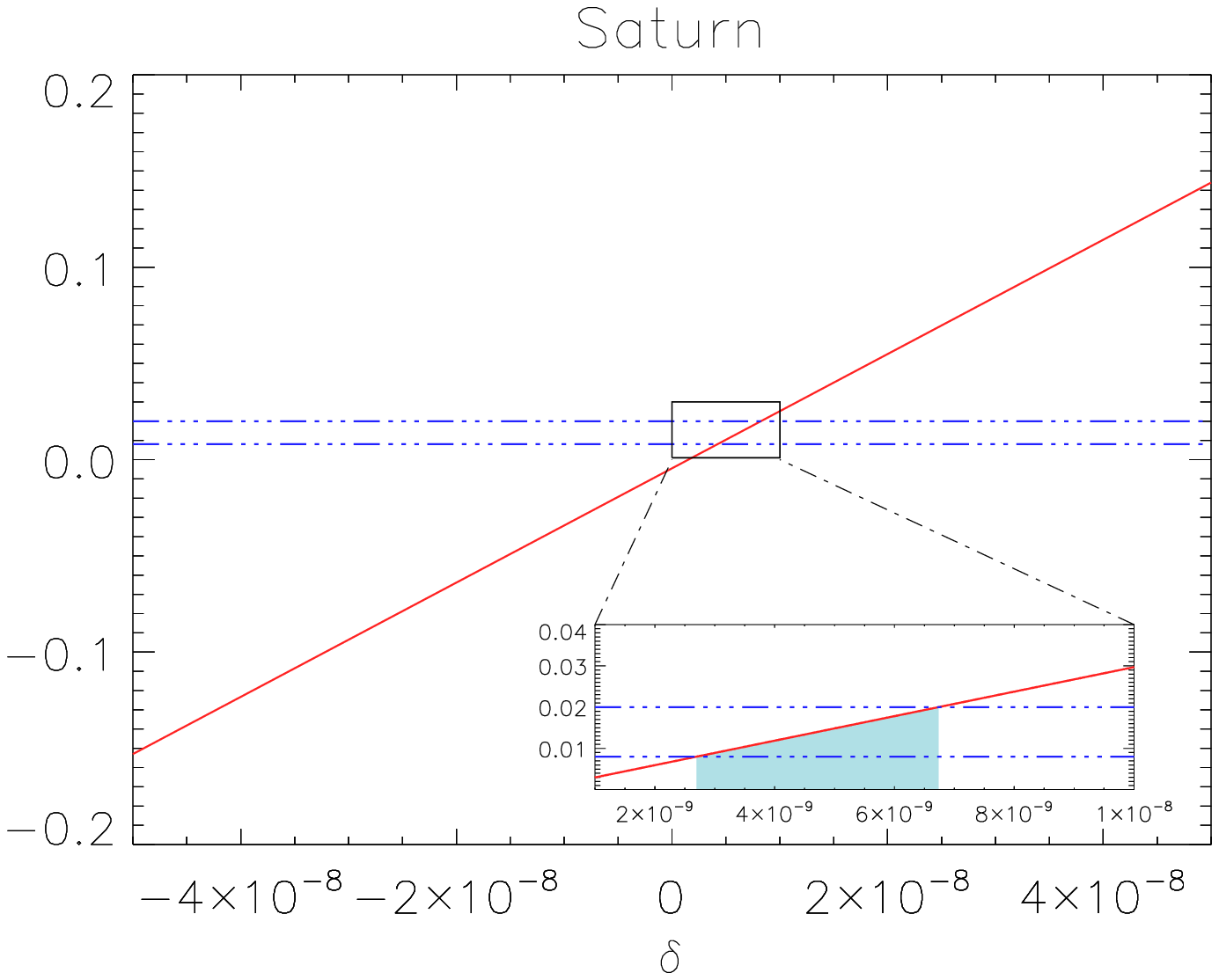}\\
\caption{Planetary precession in the Yukawa-like gravitational potential as a function of the force strength
$\delta$. The blue dotted lines show the 99\% confidence level (C.L.) of the measurements; the red lines give the theoretical 
prediction as in Eq. \eqref{eq:precYuk}. The shaded zones depict the allowed range of the strength parameter.}\label{fig4}
\end{center}
\end{figure*}

Finally, defining the auxiliary variable $\xi\equiv(1+\delta)\lambda^2$, we find
\begin{widetext}
\begin{align}\label{eq:precYuk}
\Delta \varphi &= \pi\sqrt{1+\frac{2 \delta }{-3 \delta +2 u_0^2 \xi }} \biggl(1 -\frac{2 u_0  \delta  \xi}{3 \delta ^2-8 u_0^2 \delta  \xi +4 u_0^4 \xi^2} \eta 
 + \frac{  \delta   \left(-3 \delta ^3+16 u_0^2 \delta ^2 \xi -124 u_0^4 \delta  \xi^2+144 u_0^6 \xi ^3\right)}{16 \left(3 u_0 \delta ^2-8 u_0^3 \delta \xi +4 u_0^5 \xi^2\right)^2}\eta ^2\biggr)
\end{align}
\end{widetext}
To bound the strength $\delta$, we use the motion of the Solar system's planets. 
Specifically, we use Mercury, Venus, Earth, Mars, Jupiter and Saturn 
for which the orbital precession has been measured \cite{Nyambuya2010}. 
We identify the allowed  region of $\delta$ for which the predicted precession does not contradict the data.
In Figure \ref{fig4} we show the allowed zone of parameter space for each planet (light blue shades), and 
we also show that for $-1<\delta<0$ the precession in ongoing in the opposite direction with respect
the observed one, while  $\delta>0$ give rise to a precession in the right direction confirming 
the results found for $R^n$ gravity using the S-stars orbiting around the Galactic Center \cite{Borka20012, Borka20013}. 
This results was rather expected since the effect produced by the modification of the gravitational potential must be 
greater or lower than the Newtonian one that is zero. 
The scale length has been fixed to the confidence value $\lambda=5000$ AU \cite{Zakharov2016}, and a lower and upper
limit on $\delta$ is inferred, and reported in Table \ref{tab:1}. The tightest interval on $\delta$ is obtained
with Saturn that is located at the highest distance from the Sun. This restricts $\delta$ to vary in the range from 
$2.70\times10^{-9}$ to $6.70\times10^{-9}$. With these values of the strength we have also predicted the precession for 
Uranus, Neptune and Pluto\footnote{Although the latter is not a planet, its large distance from the Sun and its small mass makes
the object very useful to show the impact of the modified gravitational potential.}, and we found a precession up to three
order of magnitude larger than the one predicted by General Relativity meaning that the strength must be even smaller than 
$fews \times 10^{-9}$ to match the general relativistic constraints. All results are summarized in Table \ref{tab:2}.
\begin{table*}
\caption{\label{tab:1} This table reports for different planets 
the observed values of the: semi-major axis ($a$), orbital period ($P$), tilt angle ($i$),
eccentricity ($e$), orbital precession in columns $2-6$, respectively. In column $8$ 
we give the predicted precession in General Relativity, and finally in the last 
column we give the bounds on $\delta$. Currently no data are available for the other Solar System planets.
The table has been adapted from \cite{Nyambuya2010}.}
\begin{tabular}{l c c c c c c c}
\hline\hline\\
\multicolumn{5}{c}{}&\multicolumn{3}{c}{\textbf{Precession}\,\,\,($1\prime\prime/100{yrs}$)}\\
\multicolumn{5}{c}{}&\multicolumn{3}{c}{\textbf{{---------------------------------------------------------------------}}}\\
\textbf{Planet} & \multicolumn{1}{c}{$a$} & \multicolumn{1}{c}{$ P $} & $i$  & $\epsilon$ &
$\dot{\omega}_{obs}$ & $\dot{\omega}_{GR}$ & $[\delta_{min}; \delta_{max}]$ \\
 & ({AU})  & ({yrs}) & (degrees)  & &  \\
\hline\\
{\textbf{Mercury}}  & ${0.39}$ & ${0.24}$   & ${7.0}$  & ${0.206}$ & $43.1000\pm0.5000$ & $43.5$          & $[1.02;\, 1.09]\times10^{-2}$ \\
{\textbf{Venus}}    & ${0.72}$ & ${0.62}$   & ${3.4}$  & ${0.007}$ & $8.0000\pm5.0000$  & $\,\,\,8.62$    & $[-0.76;\, 2.51]\times10^{-3}$\\
{\textbf{Earth}}    & ${1.00}$ & ${1.00}$   & ${0.0}$  & ${0.017}$ & $5.0000\pm1.0000$  & $\,\,\,3.87$    & $[1.45;\, 5.79]\times10^{-4}$\\
{\textbf{Mars}}     & ${1.52}$ & ${1.88}$   & ${1.9}$  & ${0.093}$ & $1.3624\pm0.0005$ & $\,\,\,1.36$     & $[5.90;\, 5.92]\times10^{-5}$\\
{\textbf{Jupiter}}  & ${5.20}$ & ${11.86}$  & ${1.3}$  & ${0.048}$ & $0.0700\pm0.0040$  & $\,\,\,0.0628$  & $[0.92;\, 1.30]\times10^{-7}$\\
{\textbf{Saturn}}   & ${9.54}$ & ${29.46}$  & ${2.5}$  & ${0.056}$ & $0.0140\pm0.0020$  & $\,\,\,0.0138$  & $[2.70;\, 6.70]\times10^{-9}$\\
\hline\hline\\
\end{tabular}\\
\end{table*}

\begin{table*}
\caption{\label{tab:2} This table reports observations from \cite{Nyambuya2010} as well.
The last two columns report the prediction of the precession from General Relativity and from the Yukawa-like gravitational
potential. The latter have been computed using the tight bounds of $\delta$ from Saturn (see Table \ref{tab:1}) and fixing the 
characteristic length to $\lambda=5000$ AU.}
\begin{tabular}{l c c c c c c}
\hline\hline\\
\multicolumn{5}{c}{}&\multicolumn{2}{c}{\textbf{Precession}\,\,\,($1\prime\prime/100{yrs}$)}\\
\multicolumn{5}{c}{}&\multicolumn{2}{c}{\textbf{{-------------------------------------}}}\\
\textbf{Planet} & \multicolumn{1}{c}{$a$} & \multicolumn{1}{c}{$ P $} & $i$  & $\epsilon$ &
$\dot{\omega}_{GR}$ &  $[\dot{\omega}_{min}; \dot{\omega}_{max}]|$\\
 & ({AU})  & ({yrs}) & (degrees)  &  \\
\hline\\
{\textbf{Uranus}}   & ${19.2}$ & ${84.10}$  & ${0.8}$  & ${0.046}$ &  $\,\,\,0.0024$  & $[0.05;\,0.12]$\\
{\textbf{Neptune}}  & ${30.1}$ & ${164.80}$ & ${1.8}$  & ${0.009}$ &  $\,\,\,0.00078$ & $[0.18;\,0.45]$\\
{\textbf{Pluto}}    & ${39.4}$ & ${247.70}$ & ${17.2}$ & ${0.250}$ &  $\,\,\,0.00042$ & $[0.11;\,0.30]$\\
\hline\hline\\
\end{tabular}\\
\end{table*}

\section{Implications for $f(R)$ gravity}
\label{cinque}

To make compatible $f(R)$ models with local gravity constraints, these theories usually
require a "screening mechanism". When considering theories with a non-minimally coupled
scalar field, one has to impose strong conditions on the effective mass of the scalar field that 
must depend on the space-time curvature or, alternatively, on the matter density distribution of 
the environment \cite{Khoury2004, Khoury2009}. Thus, the scalar field can have a short 
range 
at Solar System scale 
escaping the experimental constraints, 
and have a long range at the cosmological scale, 
where it can propagate freely affecting the cosmological dynamics, and driving  the  accelerated 
expansion (see for details \cite{defelice2010}).
With the same aim, similar mechanisms have been proposed for other models, such as the symmetron and the braneworld 
\cite{Dvali2000, Nicolis2009, Hinterbichler2010}. 
{ Nevertheless, these mechanisms are introduced {\it ad hoc} and particularized for each theory. In $f(R)$ gravity, 
the need of introducing a screening mechanism arises when, instead of working with higher order field equations, one 
performs a conformal transformation from the Jordan to the Einstein frame, where the field equations are of 
second order but a scalar field, related to the $f'(R)$ term, appears. Although it is 
simpler to work with second order field equations, and the two frames are mathematically equivalent, 
one should remember that the physical equivalence is not guaranteed in general \cite{Magnano,Faraoni,darkmetric}. Thus, one could prefer to 
work with high order field equations, staying in the Jordan frame, and handling the extra degrees of freedom as 
free parameters to be constrained by the data. In such a case, the scale dependence of these parameters plays the role of 
the screening mechanism. The screening mechanism is traced by the density of the self gravitating systems \cite{chameleon}.} 

\begin{figure}
\centering
\includegraphics[width=8.6cm]{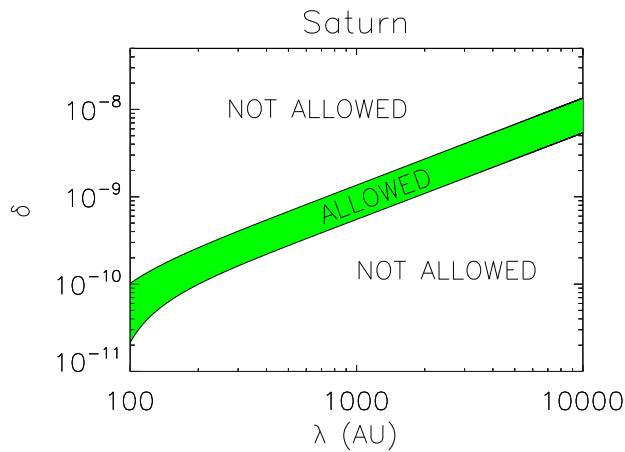}
\caption{Degeneracy between the strength and the scale length of the Yukawa gravitational potential. The plot is particularized
for the case of Saturn.}\label{fig5}
\end{figure}

Relatively, the results in Table \ref{tab:1} can be straightforwardly interpreted as the fact that the Yukawa correction 
term to the Newtonian gravitational potential is screened at planetary scales. Indeed, the departure from Newtonian 
gravity is of the order of $10^{-9}$ in $\delta$. 
Finally, the values of the strength and the scale of the Yukawa potential highly degenerate at such small scales. 
To illustrate this degeneracy we have computed the lower and upper limit on $\delta$ varying $\lambda$ from 100 AU to $10^{4}$ AU. 
The results are shown in Fig. \ref{fig5}, where we have highlighted the parts of the parameter space that are (and are not) allowed.
We show that a change of one order of magnitude in the scale length is reflected in change up to two order of magnitude in $\delta$.
The plot is particularized for Saturn.
\section{Conclusions and Remarks}
\label{sei}

Measurements of the orbital precession of Solar System bodies can be used to compare observations
with theoretical predictions arising from alternative theories of gravity. Specifically, $f(R)$ gravity models
that, in their weak field limit, show a Yukawa-like correction to the Newtonian gravitational potential can be 
used to compute the orbital precession  with a classical mechanics approach. We have computed an analytical expression
for the orbital precession and compared its prediction with the values for the Solar System's planets. 
We found that, fixing the characteristic
scale length to $\lambda=5000$ AU \cite{Zakharov2016}, the strength must rely in the range $[2.70; 6.70]\times10^{-9}$. 
Nevertheless, we must point out the presence of a  degeneracy between the strength and the scale of the Yukawa potential.
We find the direction of the orbital precession changing 
with the sign of the strength, confirming previous results \cite{Borka20012, Borka20013}. 
If the change of the direction of the orbital precession can be
used as an effective way to discriminate between General Relativity and its alternative, 
should be studied in a full relativistic approach 
where the motion happens along the geodesics \cite{pII}.

\section*{Acknowledgements}
I.D.M acknowledge financial supports from University of the Basque Country UPV/EHU under the program
"Convocatoria de contrataci\'{o}n para la especializaci\'{o}n de personal 
investigador doctor en la UPV/EHU 2015", from the Spanish Ministerio de
Econom\'{\i}a y Competitividad through the research project FIS2017-85076-P (MINECO/AEI/FEDER, UE),
and from  the Basque Government through the research project IT-956-16. 
M.\,D.\,L.\ is supported by the ERC Synergy Grant
``BlackHoleCam'' -- Imaging the Event Horizon of Black Holes (Grant
No.~610058). M.D.L. acknowledge INFN Sez. di Napoli (Iniziative Specifiche QGSKY and TEONGRAV).
This article is based upon work from COST Action CA1511 Cosmology and Astrophysics 
Network for Theoretical Advances and Training Actions (CANTATA), 
supported by COST (European Cooperation in Science and Technology).



\begin{thebibliography}{99}

\bibitem{Planck16_13}
Planck Collaboration, \aea \, 594, A13 (2016).

\bibitem{Bertone2005}
G. Bertone, D. Hooper, J. Silk, 
Phys. Rept.,  405, 279 (2005)


\bibitem{Capolupo2010}
A. Capolupo,  Advances in High Energy Physics, 2016, 8089142 (2010)

  \bibitem{Schive2014}
Hsi-Yu Schive, T. Chiueh, T. Broadhurst, 
Nature Physics, 10,  7, 496-499 (2014)

\bibitem{demartino2017b}
I. De Martino, T. Broadhurst, S.-H. H. Tye, T. Chiueh, Hsi-Yu Schive, R. Lazkoz,
\prl, 119, 221103 (2017)

\bibitem{demartino2018}
I. De Martino, T. Broadhurst, S.-H. H. Tye, T. Chiueh, Hsi-Yu Schive, R. Lazkoz, 
Galaxies, 6, 10 (2018)

\bibitem{Lopes2018}
I. Lopes, G. Panotopoulos
(2018) arXiv:1801.05031 

\bibitem{Panotopoulos2018}
G. Panotopoulos, I. Lopes, 
(2018) arXiv:1801.03387

\bibitem{Feng2010}
J.L. Feng, Ann. Rev. Astron. Astrophys. 48, 495, (2010).

\bibitem{Moffat2006}
J. W. Moffat, J. Cosmol. Astropart. Phys. 0603, 004 (2006).

\bibitem{demartino2017}
I. de Martino, M. De Laraurentis,
\plb \, {770}, 440 (2017)

\bibitem{PhysRept}
S. Capozziello, M. De Laurentis,  Phys. Rept. 509, 167 (2011).

\bibitem{manos}
Y.-F. Cai, S. Capozziello, M. De Laurentis, E. N. Saridakis, Rept. Prog. Phys. {\bf 79}, 106901 (2016).

\bibitem{sergei}
S.~Nojiri and S.~D.~Odintsov,
Phys.\ Rept.\  {\bf 505}, 59 (2011) 

\bibitem{Bogdanos:2009tn}
C.~Bogdanos, S.~Capozziello, M.~De Laurentis and S.~Nesseris,
Astropart.\ Phys.\  {\bf 34} 236 (2010). 

\bibitem{felix}M.~De Laurentis, O.~Porth, L.~Bovard, B.~Ahmedov and A.~Abdujabbarov,
  Phys.\ Rev.\ D {\bf 94} no.12,  124038 (2016). 


\bibitem{Bellucci:2008jt}
S.~Bellucci, S.~Capozziello, M.~De Laurentis and V.~Faraoni,
Phys.\ Rev.\ D {\bf 79}, 104004 (2009) 

\bibitem{graviton}S. Capozziello, M. De Laurentis, S.Nojiri, S.D. Odintsov,Phys. Rev. D {\bf 95} 083524 (2017)

\bibitem{Annalen} 
S. Capozziello, M. De Laurentis, Annalen der Physik 524, 545 (2012).

\bibitem{Quandt1991}
 I. Quandt, H. J. Schmidt, Astron. Nachr., 312, 97 (1991).

\bibitem{Lee2010}
K. Lee, F. A. Jenet, R. H. Price, N. Wex, M. Kramer
\apj \, 722, 1589-1597 (2010) 
 
\bibitem{Abbott2017}
B. P. Abbott et al. (LIGO Scientific and Virgo Collaboration)
\prl \, 118, 221101 (2017)

\bibitem{Cap-def-Sal2009}
S. Capozziello, E. De Filippis, V.Salzano,  \mnras {\bf 394}, 947 (2009)
 
\bibitem{demartino2014}
I. de Martino, M. De Laurentis, F. Atrio-Barandela, S. Capozziello,
\mnras \, 442, 2, 921-928 (2014)

\bibitem{demartino2015}
I. de Martino, M. De Laurentis, S. Capozziello,
Universe 1, 123 (2015)

\bibitem{demartino2016}
I. de Martino, \prd \,  93, 124043 (2016)

\bibitem{Talmadge1988}
C. Talmadge, J.-P. Berthias, R. W. Hellings and E. M. Standish,
\prl \, 61, 1159 (1988).

\bibitem{Anderson1998}
J. D. Anderson, P. A. Laing, E. L. Lau, A. S. Liu, M. M. Nieto, and S. G. Turyshev, 
\prl \, 81, 2858 (1998).

\bibitem{Anderson2002}
J. D. Anderson, P. A. Laing, E. L. Lau, A. S. Liu, M. M. Nieto, and S. G. Turyshev, 
\prd \, 65, 082004 (2002).

\bibitem{Borka20012}
D. Borka, P. Jovanovi\'c, V. Borka Jovanovi\'c and A. F. Zakharov,
 \prd \, 85 124004 (2012)

\bibitem{Borka20013}
D. Borka, P. Jovanovi\'c, V. Borka Jovanovi\'c, A. F. Zakharov,
\jcap \, 11 050 (2013).

\bibitem{Zakharov2016}
A.F. Zakharov and P. Jovanovi\'c and D. Borka and V. Borka Jovanovi\'c,
\jcap \, 05,045, (2016) 

\bibitem{Hees2017}
A. Hees, T. Do, A. M. Ghez, G. D. Martinez, S. Naoz, E. E. Becklin, A. Boehle, S. Chappell, 
D. Chu, A. Dehghanfar, et al.
\prl 118, 211101, (2017)

\bibitem{Zakharov2018}
A. F. Zakharov, P. Jovanovi\'{c}, D. Borka, V. Borka Jovanovi\'{c},  (2018)  	arXiv:1801.04679 

\bibitem{Iorio2005}
L. Iorio, \cqg, 22, 5271 (2005); L. Iorio, High Energy Physics, 2012, 73, (2012); L. Iorio, Phys. Lett. A, 298, 315 (2002);L. Iorio, Planetary and Space Science, 55, 1290 (2007).

\bibitem{deLa_deMa2014}
M. De Laurentis, I. de Martino,
\mnras \, 431, 1, 741 (2013)

\bibitem{deLa_deMa2015}
M. De Laurentis, I. de Martino,
Int. J. Goem. Math. Meth., 1250040D (2015)

\bibitem{LeeS2017}
S. Lee, (2017) arXiv:1711.09038.

\bibitem{Will2014}
C. Will, Living Reviews in Relativity, 17 (2014)

\bibitem{Iorio2009}
 L. Iorio, Astron. J. \, 137, 3 (2009) 
 
\bibitem{islam1983}
J. N. Islam, Phys. Lett. A 97, 239 (1983).

\bibitem{Iorio2006}
L. Iorio, Int. J. Mod. Phys. D 15, 473 (2006).

\bibitem{Sereno2006}
M. Sereno and P. Jetzer, \prd \, 73, 063004 (2006).

\bibitem{Gron1996}
\O{}. Gr\o{}n and H. H. Soleng,  \apj \, 456, 445 (1996).

\bibitem{Khriplovich2006}
I. B. Khriplovich and E. V. Pitjeva, Int. J. Mod. Phys. D 15 , 615 (2006).

\bibitem{Capozziello2001}
S. Capozziello, S. De Martino, S. De Siena, and F. Illuminati, Mod. Phys. Lett. A 16, 693 (2001).


\bibitem{Sanders2006}
R. H. Sanders, \mnras \, 370 , 1519 (2006).

 \bibitem{Battat2008}
J. B. R. Battat, C. W. Stubbs, J. F. Chandler, 
 \prd \, 78, 022003, (2008)

\bibitem{Nyambuya2010}
 G. G. Nyambuya,  \mnras \, 403, 1381 (2010) 

\bibitem{Ozer2017}
H. \"{O}zer, \"{O}. Delice,  	arXiv:1708.05900 (2017)

\bibitem{Liu2018}
Tan Liu, Xing Zhang, Wen Zhao
\plb B 777  286-293 (2018)

\bibitem{deLa2011}
M. De Laurentis, The Open Astronomy Journal, 4, 108-150  (2011)

\bibitem{Khoury2004}
 J. Khoury, and A. Weltman,  \prl \, 93, 171104, (2004).

\bibitem{Khoury2009}
 J. Khoury, and M. Wyman, \prd \, 80, 064023,  (2009).


\bibitem{defelice2010}
A. de Felice, and S. Tsujikawa, Living Reviews in Relativity, 13, 3, (2010).


\bibitem{Dvali2000}
G. Dvali, G. Gabadadze, M. Porrati, \plb \, 485, 208-214, (2000).

\bibitem{Nicolis2009}
A. Nicolis, R. Rattazzi, E. Trincherini, \prd \, 79, 064036, (2009).


\bibitem{Hinterbichler2010}
K. Hinterbichler, J. Khoury ,  \prl \,, 104, 231301, (2010).

\bibitem{pII}
M. De Laurentis, I. De Martino, R. Lazkoz, arXiv:1801.08136 (2017).

\bibitem{Magnano}G. Magnano, L. Sokolowski, Phys. Rev. D, {\bf 50} 5039, (1994)
\bibitem{Faraoni}V. Faraoni, E. Gunzig, Int.J.Theor.Phys., 38:217?225 (1999)
\bibitem{darkmetric}
S. Capozziello, M. De Laurentis, M. Francaviglia, S. Mercadante, Foundations of Physics {\bf 39}, 1161 (2009).
\bibitem{chameleon}S. Capozziello, S. Tsujikawa , Phys. Rev. D, {\bf  77},107501 (2008).


\end{thebibliography}
\end{document}